\documentclass{appolb}
\usepackage{graphicx}
\usepackage{slashed}
\usepackage{multirow}
\usepackage{subcaption}
\usepackage{rotating}
\usepackage[table]{xcolor}
\usepackage{enumitem}
\usepackage{colortbl}
\usepackage{pdflscape}
\usepackage{float}
\usepackage{siunitx}
\usepackage[compat=1.1.0]{tikz-feynman}
\usepackage{lscape}
\usepackage{mathtools}
\usepackage{amsfonts}
\usepackage{bbold}
\usepackage{soul}
\usepackage{comment}
\usepackage{adjustbox}
\usepackage{lineno}
\usepackage{tocloft}
\usepackage[compress,numbers,sort]{natbib}

\newcommand{\squa}[3]{\ensuremath{{[#1|}_{#3}^{#2}}}
\newcommand{\squet}[3]{\ensuremath{|#1]_{#3}^{#2}}}

\newcommand{\tret}[3]{\ensuremath{{|#1\rangle}_{#3}^{#2}}}

\newcommand{\squaket}[2]{\ensuremath{[#1|#2]}}

\begin{document}
\title{Recent developments in SMEFT:\\
theory, tools and phenomenology%
\thanks{Presented at the XLVI International Conference of Theoretical Physics “Matter to
the Deepest”, Katowice, Poland, 15–19 September, 2025}%
}
\author{Michał Ryczkowski
\address{Dipartimento di Fisica e Astronomia ``G. Galilei'', Universit\`a di Padova and Istituto Nazionale di Fisica Nucleare, Sezione di Padova, Via F. Marzolo 8, I-35131, Padova, Italy}
\\[3mm]
}
\maketitle
\begin{abstract}
Despite the remarkable success of the Standard Model in describing fundamental interactions, unresolved phenomena such as dark matter, dark energy, and matter-antimatter asymmetry strongly suggest the existence of physics beyond the Standard Model. The absence of new particle discoveries at the LHC indicates that such new physics may be significantly heavier than the electroweak scale. In this context, Effective Field Theories offer a powerful framework for studying the indirect effects of heavy new physics. This contribution reviews some of the recent advancements, computational tools, and phenomenology of Effective Field Theories, with a particular focus on the Standard Model Effective Field Theory.
\end{abstract}

\newpage

\section{Introduction}
 
 The Standard Model (SM) of particle physics, developed and refined throughout the XXth century, reached its culmination with the discovery of the Higgs boson in 2012 by the ATLAS and CMS collaborations~\cite{ATLAS:2012yve,CMS:2012qbp}. While the SM provides an exceptionally precise description of fundamental interactions, it leaves several critical questions unanswered, such as the nature of dark matter and dark energy, the origin of matter-antimatter asymmetry, and the hierarchy problem. These unresolved issues strongly motivate the search for physics beyond the Standard Model (BSM).

Despite extensive searches, no direct evidence of new particles has been observed at the LHC, suggesting that the scale of new physics may lie well above the electroweak scale. In this context, Effective Field Theories (EFT) offer a powerful and systematic framework for indirectly probing BSM physics. Two prominent EFT frameworks, the Standard Model Effective Field Theory (SMEFT)~\cite{Buchmuller:1985jz,Grzadkowski:2010es} and the Higgs Effective Field Theory (HEFT)~\cite{Feruglio:1992wf,Buchalla:2012qq,Alonso:2012px}, have emerged as key tools for indirect probes of BSM phenomena. 

This contribution reviews some of the recent advancements in the theory, tools, and phenomenology of EFT, with a focus on SMEFT.
Section~\ref{sec:EFT:BSM} introduces SMEFT and HEFT, emphasizing their complementary roles in probing BSM physics. Section~\ref{sec:SMEFT:pheno} outlines examples of phenomenological applications of SMEFT and tools developed to facilitate calculations in this framework. Finally, Section~\ref{sec:onshell} explores recent progress in on-shell methods and their application to EFT.

\section{Effective Field Theories for BSM physics}
\label{sec:EFT:BSM}

\subsection{SMEFT}
The Standard Model Effective Field Theory provides a model-independent framework for parameterizing the effects of heavy and decoupled BSM physics. By assuming a linear realization of electroweak symmetry breaking (EWSB) and embedding the Higgs boson within an $SU(2)_L$ doublet, SMEFT allows for a systematic expansion in terms of operator mass dimensions. This framework incorporates the effects of new physics through higher-dimensional operators, which are suppressed by powers of a heavy new physics scale $\Lambda$. The SMEFT Lagrangian can be expressed as an expansion in $1/\Lambda$:
\begin{equation}
\mathcal{L}_{\mathrm{SMEFT}}=\mathcal{L}_{\mathrm{SM}}+
\sum_i
\frac{C_i}{\Lambda^{d_i-4}}\mathcal{O}_i \, ,
  \label{eq:SMEFT:Lagr:General}
\end{equation}
where $\mathcal{L}_{\mathrm{SM}}$ is the Standard Model Lagrangian, $\mathcal{O}_i$ are higher-dimensional operators of dimension $d_i > 4$, parametrized by dimensionless Wilson coefficients (WC) $C_i$ and invariant under the SM gauge group. For a comprehensive review, see for example~\cite{Brivio:2017vri,Isidori:2023pyp, Aebischer:2025qhh}

In recent years, SMEFT has become one of the primary tools for new physics searches with precision measurements. However, it is not universally applicable. Specifically, two types of models have been identified where more general Higgs Effective Field Theory is the appropriate framework~\cite{Banta:2021dek}: i) models in which new particles derive most of their mass from EWSB, and ii) models featuring additional sources of EWSB even in the limit of $v \to 0$, where $v$ denotes the SM vacuum expectation value.

\subsection{HEFT}
In contrast to SMEFT, HEFT treats the Higgs boson as a gauge singlet, distinct from the Goldstone bosons associated with EWSB. This distinction enables HEFT to describe a broader range of BSM scenarios, including those with non-linear EWSB, such as composite Higgs models. Unlike SMEFT, the effective Lagrangian in HEFT cannot be organized as a conventional expansion in operator mass dimensions. Instead, HEFT operators may include arbitrary powers of the dimensionless ratio $h/v$, where $v$ is the vacuum expectation value. The power counting in HEFT is more intricate and governed by the chiral dimension, reflecting the momentum and loop expansion characteristic of non-linear EFT~\cite{Buchalla:2013eza, Gavela:2016bzc, Brivio:2025yrr}. This approach facilitates a decoupling of the Higgs dynamics from the symmetry-breaking pattern, allowing for more general deviations in Higgs couplings. Such deviations must be independently constrained through precision measurements.

\subsubsection{SMEFT operator bases}
Over the past 15 years, significant progress has been made in constructing complete and non-redundant operator bases for SMEFT. The most relevant contributions arise from dimension-6 operators, (with the single dimension-5 operator), with the \textit{Warsaw basis}~\cite{Grzadkowski:2010es} serving as the most widely adopted basis for dimension-6 SMEFT. It enables consistent interpretations of precision measurements and collider data.

Bases for higher-dimensional operators have also been constructed, including dimension-7~\cite{Lehman:2014jma}, dimension-8 operators~\cite{Murphy:2020rsh,Li:2020gnx}, dimension-9 operators~\cite{Liao:2020jmn}, and beyond (up to dimension-12~\cite{Harlander:2023psl}).

In practice, most experimental and phenomenological studies focus on dimension-6 and dimension-8 operators, as their effects are less suppressed and more accessible in current and near-future collider experiments.\footnote{Odd-dimension operators, such as those of dimension-7, violate lepton or baryon number and are thus highly constrained.}

\section{SMEFT - tools and phenomenology}
\label{sec:SMEFT:pheno}

SMEFT, while not itself a direct discovery tool, offers insights into the form and structure of the underlying BSM theory. This can be achieved through a pipeline that involves i) calculating precise predictions for various HEP processes, ii) comparing these predictions with experimental data to constrain SMEFT WC, iii) interpreting WC patterns using results from matching to UV completions, and iv) deriving insights into the underlying UV theory.

In recent years, there has been a surge in precision studies in SMEFT. This ``precision'' can be achieved in two ways. Firstly, by including higher-order corrections in perturbation theory, which has been done for processes such as gluon fusion Higgs and double-Higgs production~\cite{Deutschmann:2017qum, Alasfar:2022zyr, Heinrich:2022idm, Heinrich:2024dnz, Maltoni:2024dpn} and Higgs decays~\cite{Dedes:2018seb, Dedes:2019bew, Cullen:2020zof}. This progress has been facilitated by developments in the derivation of Renormalization Group Equations (RGE) in SMEFT at one-loop~\cite{Jenkins:2013zja, Jenkins:2013wua, Alonso:2013hga} (full results) and two-loops~\cite{DiNoi:2024ajj,Born:2024mgz,Haisch:2025lvd,Duhr:2025zqw,DiNoi:2025tka} (partial results).
 
Secondly, precision in the context of SMEFT may mean considering higher-order terms in the EFT expansion (dimension-6$^2$ and dimension-8) that can affect the results. This has been demonstrated by studies of vector boson scattering and vector boson fusion double-Higgs production processes~\cite{Dedes:2020xmo,Cappati:2022skp,Dedes:2025oda}.

Although quite universal and model-independent in parameterizing BSM phenomena, SMEFT is also a very complex framework. One can appreciate this for example by simply considering exact number of higher-dimensional operators for a given mass dimension \(d_i\)~\cite{Henning:2015alf}, as presented in Table~\ref{tab:sfr:number}.

\begin{table}[h!]
  \centering
  \begin{tabular}{|c|c|c|}
    \hline
    \(d_i\) & \(n_f=1\) & \(n_f=3\) \\
    \hline
    5    & 2     & 12 \\
    \hline
    6 & 84 & 3045 \\
    \hline
    8 & 993 & 44807 \\
    \hline
    10 & 15456 & 2092441 \\
    \hline
  \end{tabular}
  \caption{Numbers of higher-dimensional operators for a given mass dimension \(d_i\) and number of fermion generations \(n_f\), as taken from~\cite{Henning:2015alf}.}
  \label{tab:sfr:number}
\end{table}

The high number of operators in SMEFT implies significant technical complexity in theoretical calculations for physical processes and observables. In addition, already mentioned issues such as matching between UV models and SMEFT, higher-order loop contributions, RGE running, and fitting SMEFT WC to experimental data have proven numerical tools indispensable for efficient calculations and progress in SMEFT. To facilitate precision calculations in SMEFT, researchers have dedicated significant effort to develop numerical tools that address these challenges. Below, we present a selection of currently available tools, divided by categories:

\begin{itemize}
    \item \textbf{SMEFT matching to UV models and RGE running}\\ {\tt Matchete}~\cite{Fuentes-Martin:2022jrf}, {\tt Matchmakereft}~\cite{Carmona:2021xtq}, {\tt MatchingTools}~\cite{Criado:2017khh}, {\tt CoDeX}~\cite{DasBakshi:2018vni}, {\tt DsixTools}~\cite{Fuentes-Martin:2020zaz}, {\tt wilson}~\cite{Aebischer:2018bkb}, {\tt SOLD}~\cite{Guedes:2023azv}, {\tt RGESolver}~\cite{DiNoi:2022ejg},
    \item \textbf{SMEFT fitting to experimental data}\\ {\tt SMEFiT}~\cite{Giani:2023gfq}, {\tt smelli}~\cite{Stangl:2020lbh}, {\tt HepFIT}~\cite{DeBlas:2019ehy}, {\tt match2fit}~\cite{terHoeve:2023pvs},
    \item \textbf{Feynman rules and physical observables}\\ {\tt SMEFTsim}~\cite{Brivio:2017btx}, {\tt Dim6Top}~\cite{Aguilar-Saavedra:2018ksv}, {\tt SMEFT@NLO}~\cite{Degrande:2020evl}, and {\tt SmeftFR}~\cite{Dedes:2019uzs, Dedes:2023zws}.
\end{itemize}

\section{On-shell techniques and Effective Field Theory}
\label{sec:onshell}

Recent years have witnessed significant advancements in on-shell amplitude techniques within the spinor-helicity formalism (see~\cite{Elvang:2013cua} for a review) and their application beyond Quantum Chromodynamics (QCD). By leveraging fundamental principles such as Lorentz invariance, unitarity, and Bose or Fermi statistics, these techniques enable the direct construction of scattering  without explicit reference to the Lagrangian. This bootstrap approach has demonstrated considerable utility in several key areas of EFT-related research: i) constructing bases of EFT operators~\cite{Shadmi:2018xan,Aoude:2019tzn,Durieux:2019eor,AccettulliHuber:2021uoa}, ii) computing RGE running~\cite{Baratella:2020lzz,Baratella:2020dvw,Bresciani:2024shu,Aebischer:2025zxg}, and iii) matching EFT to BSM models~\cite{DeAngelis:2023bmd,Chala:2024llp}. 

Ref.~\cite{Grober:2025vse} demonstrated that the on-shell formalism, being more general than any specific EFT framework, can be effectively used to chart the differences between SMEFT and HEFT. It focused on the study of $gg\rightarrow hhh$ production in a general on-shell EFT, extending existing $gg\rightarrow h$ and $gg\rightarrow hh$ results~\cite{Shadmi:2018xan}, and included systematic matching to SMEFT and HEFT. Table~\ref{tab:amp_gghh_SMEFTHEFT} summarizes the orders of the EFT expansion at which individual contributions arise, purely from naive power counting perspective.

\renewcommand{\arraystretch}{1.25}
\begin{table}[h]
\begin{adjustbox}{width=\textwidth,max width=\textwidth,max totalheight=\textheight,keepaspectratio}
    \centering
    \begin{tabular}{c|c|c|c|c|c|c}
    \multirow{2}{*}{Amplitude} & \multirow{2}{*}{Helicity} & \multirow{2}{*}{Spinor structure} & \multirow{2}{*}{Coeff.} & \multirow{2}{*}{Dimension} & \multicolumn{2}{c}{Minimal order}  \\
    & & & & & SMEFT & HEFT \\
    \hline
    \multicolumn{7}{c}{ Three-point}\\
    \hline
    $gg\to h$     & $++$ & $\squaket{1}{2}^2$ & $c_{ggh}$ & $-1\,(1/\bar\Lambda)$ & $6\,(v/\Lambda^2)$ & $\text{NLO}^*$ \\
    $hh\to h$     & - & - & $c_{hhh}$ & $1\,(\bar\Lambda) $ & $4$ & LO \\
    \hline
    \multicolumn{7}{c}{ Four-point}\\
    \hline
    $hh\to hh$    & - & - & $c_{4h}$ & $0$ & $4$ & LO \\
    \multirow{2}{*}{$gg\to hh$}    & $++$ & $\squaket{1}{2}^2$ & $c_{gghh}^{++}$ & $-2\,(1/\bar\Lambda^2)$ & $6\,(1/\Lambda^2)$ & $ \text{NLO}^*$ \\
        & $+-$ & ${\squa{1}{}{}\mathbf{3}-\mathbf{4}\tret{2}{}{}}^2$ & $c_{gghh}^{+-}$ & $-4\,(1/\bar\Lambda^4)$ & $8\,(1/\Lambda^4)$ & $\text{NNLO}^*$ \\
    \hline
    \multicolumn{7}{c}{ Five-point}\\
    \hline
    $hh\to hhh$   & - & - & $c_{5h}$ & $0$ & $6\,(v/\Lambda^2)$ & $\text{LO}$ \\
    \hline
    \multirow{3}{*}{$gg\to hhh$}   & $++$ & $\squaket{1}{2}^2$ & $c_{gghhh}^{++,(1)}$ & $-3\,(1/\bar\Lambda^3)$ & $8\,(v/\Lambda^4)$ & $\text{NLO}^*$ \\
      & $++$ & ${\squa{1}{}{}\mathbf{3}\mathbf{4}\squet{2}{}{}}^2$ & $c_{gghhh}^{++,(2)}$ & $-7\,(1/\bar\Lambda^7)$ & $12\,(v/\Lambda^8)$ & $  \text{N$^3$LO}^*$ \\
       & $+-$ & $ {\squa{1}{}{}\mathbf{3}\tret{2}{}{}}^2$ & $c_{gghhh}^{+-}$ & $-5\,(1/\bar\Lambda^5)$ & $10\,(v/\Lambda^6)$ & $\text{NNLO}^*$ \\
    \end{tabular}
    \end{adjustbox}
    \caption{Summary of on-shell coefficients and dimensions for $gg\to hhh$, $hh\to hhh$ and the related three- and four-point amplitudes. $^*$ marks where we applied the power counting assuming that one order of $\alpha_s$ is factored out, i.e. that our HEFT operators are written as $\frac{\alpha_s}{\pi}\partial^m h^n G_{\mu\nu}G^{\mu\nu}$. }
    \label{tab:amp_gghh_SMEFTHEFT}
\end{table}

At the level of 3- and 4-point amplitudes, a naive correspondence emerges between the minimal orders at which all spinor structures appear: dimension-4 SMEFT corresponds to LO HEFT, dimension-6 SMEFT to NLO HEFT, and dimension-8 SMEFT to NNLO HEFT. However, in the case of 5-point amplitudes, things look differently. First of all, a new spinor structure, ${\squa{1}{}{}\mathbf{3}\mathbf{4}\squet{2}{}{}}^2$, emerges that was not present at lower-point amplitudes. Moreover, the naive correspondence breaks and the power counting shifts: dimension-6 SMEFT now corresponds to LO HEFT, dimension-8 SMEFT to NLO HEFT, and so on. This reveals the key distinction: SMEFT and HEFT are not fundamentally different but diverge in their convergence patterns (at least in the case of the studied processes). Future studies with heavy vector bosons and additional Higgs bosons should further clarify the relationship and differences between both frameworks.

\section{Summary}

In the absence of direct discoveries of new physics at the LHC, attention has shifted towards Effective Field Theories, particularly the Standard Model Effective Field Theory. This complex yet powerful framework required the development of numerous tools for tasks such as matching, RGE running, fitting, simulations, and matrix-element calculations. However, SMEFT has limitations in describing certain BSM scenarios, where the more general Higgs Effective Field Theory is applicable. Recently, there has been significant interest in studying the differences between SMEFT and HEFT, including the application of on-shell methods.

\bibliographystyle{utphys}
\bibliography{bibliography.bib}

@article{ATLAS:2012yve,
  title        = "{Observation of a new particle in the search for the Standard Model Higgs boson with the ATLAS detector at the LHC}",
  author       = "Aad, Georges and others",
  year         = 2012,
  journal      = "Phys. Lett. B",
  volume       = 716,
  pages        = "1--29",
  doi          = "10.1016/j.physletb.2012.08.020",
  collaboration = "ATLAS",
  eprint       = "1207.7214",
  archiveprefix = "arXiv",
  primaryclass = "hep-ex",
  reportnumber = "CERN-PH-EP-2012-218"
}

@article{CMS:2012qbp,
  title        = "{Observation of a New Boson at a Mass of 125 GeV with the CMS Experiment at the LHC}",
  author       = "Chatrchyan, Serguei and others",
  year         = 2012,
  journal      = "Phys. Lett. B",
  volume       = 716,
  pages        = "30--61",
  doi          = "10.1016/j.physletb.2012.08.021",
  collaboration = "CMS",
  eprint       = "1207.7235",
  archiveprefix = "arXiv",
  primaryclass = "hep-ex",
  reportnumber = "CMS-HIG-12-028, CERN-PH-EP-2012-220"
}

@article{Buchmuller:1985jz,
    author = "Buchmuller, W. and Wyler, D.",
    title = "{Effective Lagrangian Analysis of New Interactions and Flavor Conservation}",
    reportNumber = "CERN-TH-4254/85",
    doi = "10.1016/0550-3213(86)90262-2",
    journal = "Nucl. Phys. B",
    volume = "268",
    pages = "621--653",
    year = "1986"
}

@article{Grzadkowski:2010es,
  title        = "{Dimension-Six Terms in the Standard Model Lagrangian}",
  author       = "Grzadkowski, B. and Iskrzynski, M. and Misiak, M. and Rosiek, J.",
  year         = 2010,
  journal      = "JHEP",
  volume       = 10,
  pages        = "085",
  doi          = "10.1007/JHEP10(2010)085",
  eprint       = "1008.4884",
  archiveprefix = "arXiv",
  primaryclass = "hep-ph",
  reportnumber = "IFT-9-2010, TTP10-35"
}

@article{Feruglio:1992wf,
    author = "Feruglio, F.",
    title = "{The Chiral approach to the electroweak interactions}",
    eprint = "hep-ph/9301281",
    archivePrefix = "arXiv",
    reportNumber = "DFPD-92-TH-50",
    doi = "10.1142/S0217751X93001946",
    journal = "Int. J. Mod. Phys. A",
    volume = "8",
    pages = "4937--4972",
    year = "1993"
}

@article{Buchalla:2012qq,
    author = "Buchalla, Gerhard and Cata, Oscar",
    title = "{Effective Theory of a Dynamically Broken Electroweak Standard Model at NLO}",
    eprint = "1203.6510",
    archivePrefix = "arXiv",
    primaryClass = "hep-ph",
    reportNumber = "LMU-ASC-19-12, FLAVOUR-267104-ERC-9, LMU-ASC\textasciitilde{}19-12",
    doi = "10.1007/JHEP07(2012)101",
    journal = "JHEP",
    volume = "07",
    pages = "101",
    year = "2012"
}

@article{Alonso:2012px,
    author = "Alonso, R. and Gavela, M. B. and Merlo, L. and Rigolin, S. and Yepes, J.",
    title = "{The Effective Chiral Lagrangian for a Light Dynamical ''Higgs Particle''}",
    eprint = "1212.3305",
    archivePrefix = "arXiv",
    primaryClass = "hep-ph",
    reportNumber = "FTUAM-12-115, IFT-UAM-CSIC-12-113, CERN-PH-TH-2012-335, DFPD-2012-TH-23",
    doi = "10.1016/j.physletb.2013.04.037",
    journal = "Phys. Lett. B",
    volume = "722",
    pages = "330--335",
    year = "2013",
    note = "[Erratum: Phys.Lett.B 726, 926 (2013)]"
}

@article{Banta:2021dek,
    author = "Banta, Ian and Cohen, Timothy and Craig, Nathaniel and Lu, Xiaochuan and Sutherland, Dave",
    title = "{Non-decoupling new particles}",
    eprint = "2110.02967",
    archivePrefix = "arXiv",
    primaryClass = "hep-ph",
    doi = "10.1007/JHEP02(2022)029",
    journal = "JHEP",
    volume = "02",
    pages = "029",
    year = "2022"
}

@article{Brivio:2017vri,
    author = "Brivio, Ilaria and Trott, Michael",
    title = "{The Standard Model as an Effective Field Theory}",
    eprint = "1706.08945",
    archivePrefix = "arXiv",
    primaryClass = "hep-ph",
    doi = "10.1016/j.physrep.2018.11.002",
    journal = "Phys. Rept.",
    volume = "793",
    pages = "1--98",
    year = "2019"
}

@article{Isidori:2023pyp,
    author = "Isidori, Gino and Wilsch, Felix and Wyler, Daniel",
    title = "{The standard model effective field theory at work}",
    eprint = "2303.16922",
    archivePrefix = "arXiv",
    primaryClass = "hep-ph",
    reportNumber = "ZU-TH 14/23",
    doi = "10.1103/RevModPhys.96.015006",
    journal = "Rev. Mod. Phys.",
    volume = "96",
    number = "1",
    pages = "015006",
    year = "2024"
}

@article{Henning:2015alf,
  title        = "{2, 84, 30, 993, 560, 15456, 11962, 261485, ...: Higher dimension operators in the SM EFT}",
  author       = "Henning, Brian and Lu, Xiaochuan and Melia, Tom and Murayama, Hitoshi",
  year         = 2017,
  journal      = "JHEP",
  volume       = "08",
  pages        = "016",
  doi          = "10.1007/JHEP08(2017)016",
  note         = "[Erratum: JHEP 09, 019 (2019)]",
  eprint       = "1512.03433",
  archiveprefix = "arXiv",
  primaryclass = "hep-ph",
  reportnumber = "UCB-PTH-15-14, IPMU15-0207"
}

@article{Lehman:2014jma,
    author = "Lehman, Landon",
    title = "{Extending the Standard Model Effective Field Theory with the Complete Set of Dimension-7 Operators}",
    eprint = "1410.4193",
    archivePrefix = "arXiv",
    primaryClass = "hep-ph",
    doi = "10.1103/PhysRevD.90.125023",
    journal = "Phys. Rev. D",
    volume = "90",
    number = "12",
    pages = "125023",
    year = "2014"
}

@article{Murphy:2020rsh,
  title        = "{Dimension-8 operators in the Standard Model Effective Field Theory}",
  author       = "Murphy, Christopher W.",
  year         = 2020,
  journal      = "JHEP",
  volume       = 10,
  pages        = 174,
  doi          = "10.1007/JHEP10(2020)174",
  eprint       = "2005.00059",
  archiveprefix = "arXiv",
  primaryclass = "hep-ph"
}

@article{Li:2020gnx,
  title        = "{Complete set of dimension-eight operators in the standard model effective field theory}",
  author       = "Li, Hao-Lin and Ren, Zhe and Shu, Jing and Xiao, Ming-Lei and Yu, Jiang-Hao and Zheng, Yu-Hui",
  year         = 2021,
  journal      = "Phys. Rev. D",
  volume       = 104,
  number       = 1,
  pages        = "015026",
  doi          = "10.1103/PhysRevD.104.015026",
  eprint       = "2005.00008",
  archiveprefix = "arXiv",
  primaryclass = "hep-ph"
}

@article{Liao:2020jmn,
  title        = "{An explicit construction of the dimension-9 operator basis in the standard model effective field theory}",
  author       = "Liao, Yi and Ma, Xiao-Dong",
  year         = 2020,
  journal      = "JHEP",
  volume       = "11",
  pages        = "152",
  doi          = "10.1007/JHEP11(2020)152",
  eprint       = "2007.08125",
  archiveprefix = "arXiv",
  primaryclass = "hep-ph"
}

@article{Harlander:2023psl,
    author = "Harlander, R. V. and Kempkens, T. and Schaaf, M. C.",
    title = "{Standard model effective field theory up to mass dimension 12}",
    eprint = "2305.06832",
    archivePrefix = "arXiv",
    primaryClass = "hep-ph",
    reportNumber = "TTK-22-47, TTK-22-47; P3H-23-031",
    doi = "10.1103/PhysRevD.108.055020",
    journal = "Phys. Rev. D",
    volume = "108",
    number = "5",
    pages = "055020",
    year = "2023"
}

@article{Elvang:2013cua,
    author = "Elvang, Henriette and Huang, Yu-tin",
    title = "{Scattering Amplitudes}",
    eprint = "1308.1697",
    archivePrefix = "arXiv",
    primaryClass = "hep-th",
    month = "8",
    year = "2013"
}

@article{Shadmi:2018xan,
    author = "Shadmi, Yael and Weiss, Yaniv",
    title = "{Effective Field Theory Amplitudes the On-Shell Way: Scalar and Vector Couplings to Gluons}",
    eprint = "1809.09644",
    archivePrefix = "arXiv",
    primaryClass = "hep-ph",
    doi = "10.1007/JHEP02(2019)165",
    journal = "JHEP",
    volume = "02",
    pages = "165",
    year = "2019"
}

@article{Aoude:2019tzn,
    author = "Aoude, Rafael and Machado, Camila S.",
    title = "{The Rise of SMEFT On-shell Amplitudes}",
    eprint = "1905.11433",
    archivePrefix = "arXiv",
    primaryClass = "hep-ph",
    doi = "10.1007/JHEP12(2019)058",
    journal = "JHEP",
    volume = "12",
    pages = "058",
    year = "2019"
}

@article{Durieux:2019eor,
    author = "Durieux, Gauthier and Kitahara, Teppei and Shadmi, Yael and Weiss, Yaniv",
    title = "{The electroweak effective field theory from on-shell amplitudes}",
    eprint = "1909.10551",
    archivePrefix = "arXiv",
    primaryClass = "hep-ph",
    reportNumber = "MITP/19-090",
    doi = "10.1007/JHEP01(2020)119",
    journal = "JHEP",
    volume = "01",
    pages = "119",
    year = "2020"
}

@article{AccettulliHuber:2021uoa,
    author = "Accettulli Huber, Manuel and De Angelis, Stefano",
    title = "{Standard Model EFTs via on-shell methods}",
    eprint = "2108.03669",
    archivePrefix = "arXiv",
    primaryClass = "hep-th",
    reportNumber = "QMUL-PH-21-32, SAGEX-21-17-E",
    doi = "10.1007/JHEP11(2021)221",
    journal = "JHEP",
    volume = "11",
    pages = "221",
    year = "2021"
}

@article{Baratella:2020lzz,
    author = "Baratella, Pietro and Fernandez, Clara and Pomarol, Alex",
    title = "{Renormalization of Higher-Dimensional Operators from On-shell Amplitudes}",
    eprint = "2005.07129",
    archivePrefix = "arXiv",
    primaryClass = "hep-ph",
    doi = "10.1016/j.nuclphysb.2020.115155",
    journal = "Nucl. Phys. B",
    volume = "959",
    pages = "115155",
    year = "2020"
}

@article{Baratella:2020dvw,
    author = "Baratella, Pietro and Fernandez, Clara and von Harling, Benedict and Pomarol, Alex",
    title = "{Anomalous Dimensions of Effective Theories from Partial Waves}",
    eprint = "2010.13809",
    archivePrefix = "arXiv",
    primaryClass = "hep-ph",
    reportNumber = "TUM-HEP-1291/20",
    doi = "10.1007/JHEP03(2021)287",
    journal = "JHEP",
    volume = "03",
    pages = "287",
    year = "2021"
}

@article{Bresciani:2024shu,
    author = "Bresciani, Luigi C. and Brunello, Giacomo and Levati, Gabriele and Mastrolia, Pierpaolo and Paradisi, Paride",
    title = "{Renormalization of effective field theories via on-shell methods: the case of axion-like particles}",
    eprint = "2412.04160",
    archivePrefix = "arXiv",
    primaryClass = "hep-ph",
    doi = "10.1007/JHEP10(2025)190",
    journal = "JHEP",
    volume = "10",
    pages = "190",
    year = "2025"
}

@article{Aebischer:2025zxg,
    author = "Aebischer, Jason and Bresciani, Luigi C. and Selimovic, Nudzeim",
    title = "{Anomalous dimension of a general effective gauge theory. Part I. Bosonic sector}",
    eprint = "2502.14030",
    archivePrefix = "arXiv",
    primaryClass = "hep-ph",
    reportNumber = "CERN-TH-2025-032",
    doi = "10.1007/JHEP08(2025)209",
    journal = "JHEP",
    volume = "08",
    pages = "209",
    year = "2025"
}

@article{DeAngelis:2023bmd,
    author = "De Angelis, Stefano and Durieux, Gauthier",
    title = "{EFT matching from analyticity and unitarity}",
    eprint = "2308.00035",
    archivePrefix = "arXiv",
    primaryClass = "hep-ph",
    reportNumber = "CERN-TH-2023-150",
    doi = "10.21468/SciPostPhys.16.3.071",
    journal = "SciPost Phys.",
    volume = "16",
    pages = "071",
    year = "2024"
}

@article{Chala:2024llp,
    author = "Chala, Mikael and L{\'o}pez Miras, Javier and Santiago, Jos{\'e} and Vilches, Fuensanta",
    title = "{Efficient on-shell matching}",
    eprint = "2411.12798",
    archivePrefix = "arXiv",
    primaryClass = "hep-ph",
    doi = "10.21468/SciPostPhys.18.6.185",
    journal = "SciPost Phys.",
    volume = "18",
    number = "6",
    pages = "185",
    year = "2025"
}

@article{Grober:2025vse,
    author = {Gr{\"o}ber, Ramona and Rossia, Alejo N. and Ryczkowski, Micha{\l}},
    title = "{Multi-Higgs amplitudes bootstrapped: dissecting SMEFT and HEFT}",
    eprint = "2509.02680",
    archivePrefix = "arXiv",
    primaryClass = "hep-ph",
    reportNumber = "COMETA-2025-34",
    doi = "10.1007/JHEP02(2026)245",
    journal = "JHEP",
    volume = "02",
    pages = "245",
    year = "2026"
}

@article{Deutschmann:2017qum,
    author = "Deutschmann, Nicolas and Duhr, Claude and Maltoni, Fabio and Vryonidou, Eleni",
    title = "{Gluon-fusion Higgs production in the Standard Model Effective Field Theory}",
    eprint = "1708.00460",
    archivePrefix = "arXiv",
    primaryClass = "hep-ph",
    reportNumber = "CP3-17-24, CERN-TH-2017-165, NIKHEF-2017-035",
    doi = "10.1007/JHEP12(2017)063",
    journal = "JHEP",
    volume = "12",
    pages = "063",
    year = "2017",
    note = "[Erratum: JHEP 02, 159 (2018)]"
}

@article{Alasfar:2022zyr,
    author = {Alasfar, Lina and de Blas, Jorge and Gr{\"o}ber, Ramona},
    title = "{Higgs probes of top quark contact interactions and their interplay with the Higgs self-coupling}",
    eprint = "2202.02333",
    archivePrefix = "arXiv",
    primaryClass = "hep-ph",
    reportNumber = "HU-EP-21/48-RTG",
    doi = "10.1007/JHEP05(2022)111",
    journal = "JHEP",
    volume = "05",
    pages = "111",
    year = "2022"
}

@article{Jenkins:2013zja,
    author = "Jenkins, Elizabeth E. and Manohar, Aneesh V. and Trott, Michael",
    title = "{Renormalization Group Evolution of the Standard Model Dimension Six Operators I: Formalism and lambda Dependence}",
    eprint = "1308.2627",
    archivePrefix = "arXiv",
    primaryClass = "hep-ph",
    doi = "10.1007/JHEP10(2013)087",
    journal = "JHEP",
    volume = "10",
    pages = "087",
    year = "2013"
}

@article{Jenkins:2013wua,
    author = "Jenkins, Elizabeth E. and Manohar, Aneesh V. and Trott, Michael",
    title = "{Renormalization Group Evolution of the Standard Model Dimension Six Operators II: Yukawa Dependence}",
    eprint = "1310.4838",
    archivePrefix = "arXiv",
    primaryClass = "hep-ph",
    doi = "10.1007/JHEP01(2014)035",
    journal = "JHEP",
    volume = "01",
    pages = "035",
    year = "2014"
}

@article{Alonso:2013hga,
    author = "Alonso, Rodrigo and Jenkins, Elizabeth E. and Manohar, Aneesh V. and Trott, Michael",
    title = "{Renormalization Group Evolution of the Standard Model Dimension Six Operators III: Gauge Coupling Dependence and Phenomenology}",
    eprint = "1312.2014",
    archivePrefix = "arXiv",
    primaryClass = "hep-ph",
    reportNumber = "CERN-PH-TH-2013-305",
    doi = "10.1007/JHEP04(2014)159",
    journal = "JHEP",
    volume = "04",
    pages = "159",
    year = "2014"
}

@article{DiNoi:2024ajj,
    author = {Di Noi, Stefano and Gr{\"o}ber, Ramona and Mandal, Manoj K.},
    title = "{Two-loop running effects in Higgs physics in Standard Model Effective Field Theory}",
    eprint = "2408.03252",
    archivePrefix = "arXiv",
    primaryClass = "hep-ph",
    reportNumber = "COMETA-2024-19",
    doi = "10.1007/JHEP12(2024)220",
    journal = "JHEP",
    volume = "12",
    pages = "220",
    year = "2025"
}

@article{Born:2024mgz,
    author = "Born, Lukas and Fuentes-Mart{\'\i}n, Javier and Kvedarait{\.{e}}, Sandra and Thomsen, Anders Eller",
    title = "{Two-loop running in the bosonic SMEFT using functional methods}",
    eprint = "2410.07320",
    archivePrefix = "arXiv",
    primaryClass = "hep-ph",
    doi = "10.1007/JHEP05(2025)121",
    journal = "JHEP",
    volume = "05",
    pages = "121",
    year = "2025"
}

@article{Haisch:2025lvd,
    author = "Haisch, Ulrich",
    title = "{Higgs production from anomalous gluon dynamics}",
    eprint = "2503.06249",
    archivePrefix = "arXiv",
    primaryClass = "hep-ph",
    reportNumber = "MPP-2025-38",
    doi = "10.1007/JHEP06(2025)004",
    journal = "JHEP",
    volume = "06",
    pages = "004",
    year = "2025"
}

@article{Duhr:2025zqw,
    author = "Duhr, Claude and Vasquez, Andres and Ventura, Giuseppe and Vryonidou, Eleni",
    title = "{Two-loop renormalisation of quark and gluon fields in the SMEFT}",
    eprint = "2503.01954",
    archivePrefix = "arXiv",
    primaryClass = "hep-ph",
    reportNumber = "BONN-TH-2025-07",
    doi = "10.1007/JHEP07(2025)160",
    journal = "JHEP",
    volume = "07",
    pages = "160",
    year = "2025"
}

@article{DiNoi:2025tka,
    author = {Di Noi, Stefano and Erdelyi, Barbara Anna and Gr{\"o}ber, Ramona},
    title = "{Complete two-loop Yukawa-induced running of the Higgs-gluon coupling in SMEFT}",
    eprint = "2510.14680",
    archivePrefix = "arXiv",
    primaryClass = "hep-ph",
    reportNumber = "KA-TP-31-2025, COMETA-2025-47",
    month = "10",
    year = "2025"
}

@article{Dedes:2018seb,
    author = "Dedes, A. and Paraskevas, M. and Rosiek, J. and Suxho, K. and Trifyllis, L.",
    title = "{The decay $h\to \gamma\gamma$ in the Standard-Model Effective Field Theory}",
    eprint = "1805.00302",
    archivePrefix = "arXiv",
    primaryClass = "hep-ph",
    doi = "10.1007/JHEP08(2018)103",
    journal = "JHEP",
    volume = "08",
    pages = "103",
    year = "2018"
}

\end{document}